\documentclass[iop]{emulateapj}

\begin{document}

\shorttitle{Recurrent outbursts in 3XMM J031820.8-663034}

\title{Recurrent outbursts revealed in 3XMM J031820.8-663034}

\author{Hai-Hui Zhao$^{1}$, Shan-Shan Weng$^{1}$, Jun-Xian Wang$^{2,3}$}

\affil{$^{1}$ Department of Physics and Institute of Theoretical Physics,
Nanjing Normal University, Nanjing 210023, China}

\affil{$^{2}$ CAS Key Laboratory for Research in Galaxies and Cosmology,
Department of Astronomy, University of Science and Technology of China, Hefei
230026, China}

\affil{$^{3}$ School of Astronomy and Space Science, University of Science and
Technology of China, Hefei 230026, China}

\email{wengss@njnu.edu.cn}
\begin{abstract}

3XMM J031820.8-663034, first detected by {\it ROSAT} in NGC 1313, is one of
a few known transient ultraluminous X-ray sources (ULXs). In this
paper, we present decades of X-ray data of this source from {\it ROSAT}, {\it
XMM-Newton}, {\it Chandra} and the Neil Gehrels {\it Swift} Observatory.
We find that its X-ray emission experienced four outbursts since 1992,
with a typical recurrent time $\sim$ 1800 days, an outburst duration $\sim
240-300$ days, and a nearly constant peak X-ray luminosity $\sim 1.5
\times 10^{39}$ erg/s. The upper limit of X-ray luminosity at the
quiescent state is $\sim 5.6 \times 10^{36}$ erg/s, and the total
energy radiated during one outburst is $\sim 10^{46}$ erg. The spectra at the
high luminosity states can be described with an absorbed disk
black-body, and the disk temperature increases with the X-ray luminosity. We
compare its outburst properties with other known transient ULXs including ESO
243-49 HLX-1. As its peak luminosity only marginally puts it in the category of
ULXs, we also compare it with normal transient black hole binaries. Our results
suggest that the source is powered by an accreting massive stellar-mass black
hole, and the outbursts are triggered by the thermal-viscous instability.

\end{abstract}


\section{Introduction}
Galactic low-mass X-ray binaries (LMXBs) span most their time in
quiescence and enter into outbursts occasionally with X-ray luminosity
increased by several orders of magnitude \citep[e.g. ][]{chen97,
corral16}. Black hole LMXBs manifest themselves into five well-known spectral
states \citep[see ][ for reviews]{fernder04, mcclintock06, zhang13, yuan14}. As
luminosity increases, sources leave the ``off'' state (i.e. the quiescent
state, $L_{\rm X} < 10^{34}$ erg/s), and enter the low/hard state and
the high/soft state. The X-ray spectra at the low/hard and the
quiescent states are dominated by a power-law component. In contrast,
the high/soft state is dominated by thermal emission from the
accretion disk with much weaker variabilities. Meanwhile, the very high
state and the intermediate state, representing transitions between the low/hard
state and the high/soft state \citep[e.g. ][]{mcclintock06}, have more complex
spectral and temporal behaviors. Additionally, a few rare LMXBs could
show episodes of super-Eddington accretion, e.g. V4641~Sgr
\citep{revnivtsev02} and V404~Cyg \citep[e.g. ][]{motta17}. The behaviors
become sophisticated in these outbursts, and the knowledge on them is rather
rudimentary.

It is widely accepted that outbursts of LMXBs result from
thermal-viscous instability in accretion disk \citep[e.g. ][]{chen97,
dubus01}. When the mass transfer rate from the donor star onto the compact
object is less than a critical value, the inner region of the
accretion disk is hot while the temperature of the outer disk drops
below 6000 K -- somewhere in between the disk should hence be unstable. This is
because the accretion material in this region is partially ionized, and the
hydrogen recombination results in a large change in opacity. Consequently,
this region could be out of thermal-viscous equilibrium, and the
instability would propagate in the disk, triggering an outburst. It
has been noticed that such standard disk instability model (DIM) fails
to explain some observed properties of the outbursts (e.g. the typical
``fast-rise exponential-decay'' light curves), and additional irradiation and
truncation disk effects should be taken into account \citep[see ][ for
reviews]{lasota01}.

Compared to LMXBs, ultraluminous X-ray sources (ULXs) found in nearby galaxies
are more powerful ($L_{\rm X} > 10^{39}$ erg/s) and more stable in
X-ray  \citep[see ][ for reviews]{feng11, kaaret17}. In general, the X-ray
luminosities of ULXs vary by a factor of $\leq 10$. For instance, M33
X-8 exhibits an X-ray variation amplitude of $< 50\%$ since discovered
in 1981 \citep{weng09,laparola15}. Since their temporal and spectral
properties are quite different from those observed in LMXBs, most ULXs
are suggested as massive stellar-mass black holes (MsBHs, $M_{\rm BH} \sim
20-100 M_{\odot}$) with super-Eddington accretion \citep{feng11,
weng14}. In such an system, the mass transfer rate is large enough to
keep the whole disk fully ionized (with the aid of powerful irradiation), and
therefore the accretion proceeds stably.

However, a number of ULXs show more dramatic luminosity variations,
manifesting themselves as transients \citep[e.g. ][]{Zezas06, Grimm07,
Crivellari09}. These sources may provide a bridge between LMXBs and luminosity ULXs.
Due to the lack of observations, only a few transient ULXs have been studied in
detail \citep[e.g. ][]{kaur12, middleton12}. It is yet unclear whether
the outbursts in transient ULXs are driven by the DIM.

3XMM J031820.8-663034 (R.A. = 03:18:20.8, Dec = -66:30:35) is 0.8\arcmin~away
from the center of NGC 1313. It was first detected by {\it ROSAT} with a
maximum X-ray luminosity of $L_{\rm X} \sim 1.2\times10^{39}$ erg/s
\citep{liu05} by assuming a distance of $d = 4.61$ Mpc \citep{gao15}.
Investigating a set of {\it XMM-Newton} data, \cite{lin14} found that the
source was only detected in two out of fourteen observations, revealing a
transient nature of this source. Because there are two persistent ULXs
\citep[i.e. NGC 1313 X-1 and NGC 1313 X-2, ][]{petre94, makishima00,
bachetti13, weng14, pinto2016, kosec18}, and one bright supernova \citep[SN
1978K, ][]{ryder93, zhao17} in this region, numerous X-ray observations have
been devoted to explore the spectral evolution of these sources. In this paper,
we analyze all available X-ray data of 3XMM J031820.8-663034, collected by {\it
ROSAT}, {\it Chandra}, {\it XMM-Newton}, and the Neil Gehrels {\it Swift}
Observatory, and report its recurrent activities. The data reduction and
results are described in \S2. Because its peak luminosity marginally puts it in
the category of ULXs, we compare in \S3 its outburst properties with those of
both known transient ULXs and normal LMXBs. We also discuss the origin of the
outbursts and determine the black hole mass in 3XMM J031820.8-663034 according
to the spectral investigation (\S3).

\section{Data Reduction \& Results}

\subsection{{\it XMM-Newton} data \label{xmm}}

A total of 21 {\it XMM-Newton} observation covering 3XMM J031820.8-663034 were
made from 2000 October and 2016 March (Table \ref{obs}). We reduce all data
collected from the EPIC camera \citep{struder01, turner01} using the Science
Analysis System software (\textsc{sas}) version
14.0.0\footnote{\url{https://heasarc.gsfc.nasa.gov/docs/xmm/abc/}}, and the
intervals contaminated by flaring particle background are discarded. Besides
the detections during two observations reported in \cite{lin14}, the source
turned up on MJD~56844 again. For the first two observations, only the EPIC-pn
data are analyzed because the source fell in CCD gaps of MOS1/MOS2. Due to the
same reason, we only use MOS1/MOS2 data for the observation on MJD~56844.
Circular regions with radii of 15\arcsec and 30\arcsec are adopted for the
source and background (nearby source free region), respectively. The spectral
response files are generated by the \textsc{sas} task \texttt{rmfgen} and
\texttt{arfgen}, and the spectra are grouped to have at least 15 counts per bin
with the task \texttt{specgroup} to enable the use of chi-square statistics.

\begin{figure}
\begin{center}
\includegraphics[scale=0.45]{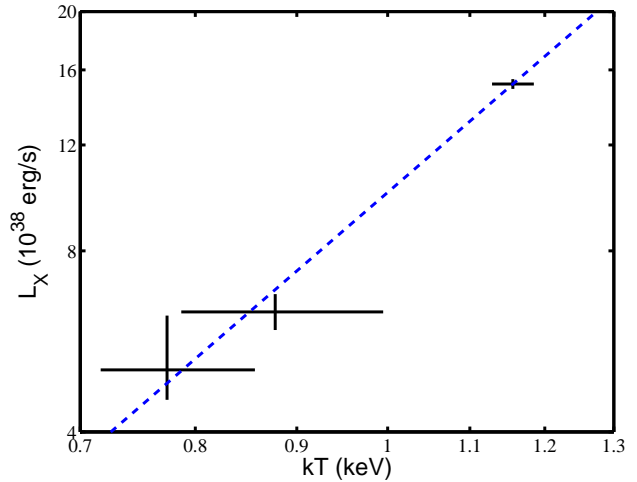}
\caption{The unabsorbed X-ray luminosity in the 0.5--10 keV band vs. disk
temperature with their 1 $\sigma$ errors. The blue dashed line
indicates the best-fit $L_{\rm X} \propto kT^{n}$ relation with $n =
2.9\pm1.3$. \label{spec}}
\end{center}
\end{figure}

Both an absorbed steep power-law ($\Gamma \sim 2.2-2.5$) and a disk black-body
\citep[{\it tbabs*diskbb} in XSPEC, ][]{arnaud96} provide adequate
fits to the first two observations (Figure \ref{xmm} and Table
\ref{fitting}). The derived parameters are consistent with those reported by
\citet[][Table 3 in their paper]{lin14}. The spectrum obtained on
MJD~56844, however, is poorly fitted by an absorbed power-law ($\chi^{2}_{\rm
\nu}/dof \sim 1.68$, Table \ref{fitting} and Figure \ref{xmm}), corresponding
to a null hypothesis probability of $1.2\times10^{-6}$. Contrarily,
the data can be well described by a disk black-body ($\chi^{2}_{\rm \nu} \sim
1.10$), and an additional power-law component is statistically not
required ($< 99\%$ according to F-$test$). As shown in Figure \ref{spec}, the
disk temperature ($kT$) increases with the unabsorbed X-ray luminosity in
0.5--10 keV. We fit the $L_{\rm X}-kT$ relation with a power-law function and
take the error of $kT$ into account. The best fitted power-law index $n =
2.9\pm1.3$ is roughly consistent with $L_{\rm X} \propto kT^{4}$ (as
predicted by the standard disk model with a constant inner radius,
i.e. the innermost stable circular orbit).

\begin{figure}
\begin{center}
\includegraphics[scale=0.55]{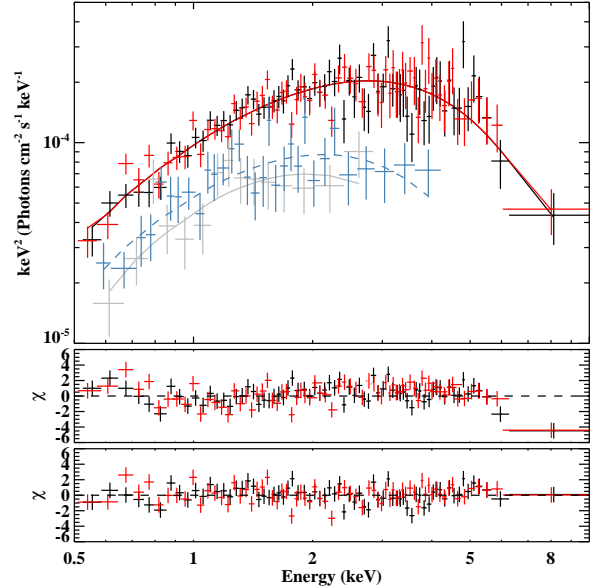}
\caption{Upper panel: {\it XMM-Newton} MOS1 (black) and MOS2 (red)
spectra observed on 2014 July 5-6 (ObsID = 0742590301), and the best-fit
absorbed $diskbb$ model. Middle and Lower panels: the fitting residuals to an
absorbed powerlaw model and an absorbed $diskbb$ model, respectively. The XMM
pn spectra observed on MJD~53332 and 53408 are also plotted in the upper panel
(grey and light-blue) for comparison. \label{xmm}}
\end{center}
\end{figure}

3XMM J031820.8-663034 was not detected in the other 18 {\it
XMM-Newton} observations, and 2$\sigma$ upper limits to the source
count rates are estimated with the \textsc{sas} task \texttt{eregionanalyse}
(Table \ref{obs}).


\begin{deluxetable*}{cccccccc}
\tabletypesize{\tiny} \tablewidth{0pt} \tablecaption{X-ray observations log}
\tablehead{\colhead{Instrument} & \colhead{ObsID} & \colhead{Date} &
\colhead{MJD} & \colhead{Energy} & \colhead{Net exposure} & \colhead{Count
rate}  &
\colhead{Flux}\\
\colhead{} & \colhead{} & \colhead{} & \colhead{} & \colhead{(keV)} &
\colhead{(ks)} & \colhead{(cts~s$^{-1}$)} & \colhead{(erg~cm$^{-2}$~s$^{-1}$)}}
\startdata

\hline

{\it XMM-Newton}/pn$^{\rm M}$  & 0106860101  & 2000-10-17 &  51834  & 0.5--10 &  22.3  & $< 3.0\times10^{-3}$ & $<1.0\times 10^{-14}$ \\
{\it XMM-Newton}/pn$^{\rm T}$  & 0150280101  & 2003-11-25 &  52968  & 0.5--10 &   1.2  & $< 1.1\times10^{-2}$ & $<3.8\times 10^{-14}$ \\
{\it XMM-Newton}/pn$^{\rm T}$  & 0150280301  & 2003-12-21 &  52994  & 0.5--10 &   8.6  & $< 4.1\times10^{-3}$ & $<1.4\times 10^{-14}$ \\
{\it XMM-Newton}/pn$^{\rm T}$  & 0150280401  & 2003-12-23 &  52996  & 0.5--10 &   4.1  & $< 3.8\times10^{-3}$ & $<1.3\times 10^{-14}$ \\
{\it XMM-Newton}/pn$^{\rm T}$  & 0150280501  & 2003-12-25 &  52998  & 0.5--10 &   6.0  & $< 8.5\times10^{-3}$ & $<2.9\times 10^{-14}$ \\
{\it XMM-Newton}/pn$^{\rm T}$  & 0150280601  & 2004-01-08 &  53012  & 0.5--10 &  10.2  & $< 4.1\times10^{-3}$ & $<1.4\times 10^{-14}$ \\
{\it XMM-Newton}/pn$^{\rm T}$  & 0150281101  & 2004-01-16 &  53020  & 0.5--10 &   5.9  & $< 6.1\times10^{-3}$ & $<2.1\times 10^{-14}$ \\
{\it XMM-Newton}/pn$^{\rm T}$  & 0205230201  & 2004-05-01 &  53126  & 0.5--10 &   0.7  & $< 1.0\times10^{-2}$ & $<3.5\times 10^{-14}$ \\
{\it XMM-Newton}/pn$^{\rm T}$  & 0205230301  & 2004-06-05 &  53161  & 0.5--10 &  10.0  & $< 6.7\times10^{-3}$ & $<2.2\times 10^{-14}$ \\
{\it XMM-Newton}/pn$^{\rm T}$  & 0205230401  & 2004-08-23 &  53240  & 0.5--10 &  10.1  & $< 4.4\times10^{-3}$ & $<1.4\times 10^{-14}$ \\
{\it XMM-Newton}/pn$^{\rm T}$  & 0205230501  & 2004-11-23 &  53332  & 0.5--10 &  12.5  & $(2.2\pm0.1)\times10^{-2}$ & $(2.0_{-0.4}^{+0.8})\times10^{-13}$ \\
{\it XMM-Newton}/pn$^{\rm T}$  & 0205230601  & 2005-02-07 &  53408  & 0.5--10 &   9.8  & $(6.2\pm0.2)\times10^{-2}$ & $(2.5_{-0.3}^{+0.3})\times10^{-13}$ \\
{\it XMM-Newton}/pn$^{\rm M}$  & 0301860101  & 2006-03-06 &  53800  & 0.5--10 &  19.9  & $< 8.5\times10^{-3}$ & $<2.9\times 10^{-14}$ \\
{\it XMM-Newton}/pn$^{\rm M}$  & 0405090101  & 2006-10-15 &  54023  & 0.5--10 &  98.8  & $< 4.3\times10^{-3}$ & $<1.5\times 10^{-14}$ \\
{\it XMM-Newton}/pn$^{\rm M}$  & 0693850501  & 2012-12-16 &  56277  & 0.5--10 & 110.7  & $< 6.3\times10^{-3}$ & $<2.1\times 10^{-14}$ \\
{\it XMM-Newton}/pn$^{\rm M}$  & 0693851201  & 2012-12-22 &  56283  & 0.5--10 & 114.8  & $< 6.2\times10^{-3}$ & $<2.1\times 10^{-14}$ \\
{\it XMM-Newton}/pn$^{\rm M}$  & 0722650101  & 2013-06-08 &  56451  & 0.5--10 &  22.3  & $< 5.3\times10^{-3}$ & $<1.8\times 10^{-14}$ \\
{\it XMM-Newton}/MOS$^{\rm M}$ & 0742590301  & 2014-07-05 &  56844  & 0.5--10 &61.0/61.0 & $(4.7\pm0.1)\times10^{-2}$ & $(5.9_{-0.2}^{+0.2})\times10^{-13}$ \\
{\it XMM-Newton}/pn$^{\rm M}$  & 0742490101  & 2015-03-30 &  57111  & 0.5--10 &  94.8  & $< 4.2\times10^{-3}$ & $<1.4\times 10^{-14}$ \\
{\it XMM-Newton}/pn$^{\rm T}$  & 0764770101  & 2015-12-05 &  57361  & 0.5--10 &  65.3  & $< 3.2\times10^{-3}$ & $<1.1\times 10^{-14}$ \\
{\it XMM-Newton}/pn$^{\rm T}$  & 0764770401 & 2016-03-23  &  57470  & 0.5--10 &  21.9  & $< 3.4\times10^{-3}$ & $<1.2\times 10^{-14}$ \\
\hline
{\it Swift}/XRT   & ...  &  Quiescence$^{*}$ & ...   & 0.3--10 &  365  & $< 2.6\times10^{-4}$ & $<1.1\times 10^{-14}$ \\
\hline
{\it ROSAT}/PSPC   & rp600045n00 & 1991-04-24 &  48371  & 0.1--2.5 & 11.0  & $< 4.9\times10^{-4}$ & $<1.7\times 10^{-14}$ \\
{\it ROSAT}/HRI    & rh400065n00 & 1992-04-18 &  48730  & 0.1--2.5 &  5.4  & $< 1.1\times10^{-3}$ & $<1.0\times 10^{-13}$ \\
{\it ROSAT}/PSPC   & rp600504n00 & 1993-11-03 &  49294  & 0.1--2.5 & 15.2  & $< 2.5\times10^{-4}$ & $<8.1\times 10^{-15}$ \\
{\it ROSAT}/HRI    & rh600505n00 & 1994-06-23 &  49527  & 0.1--2.5 & 22.6  & $(5.63\pm0.66)\times10^{-3}$ &  $5.3\times 10^{-13}$ \\
{\it ROSAT}/HRI    & rh500403n00 & 1995-01-31 &  49748  & 0.1--2.5 & 13.6  & $< 2.2\times10^{-3}$ & $<2.1\times 10^{-13}$ \\
{\it ROSAT}/HRI    & rh500404n00 & 1995-02-02 &  49750  & 0.1--2.5 & 27.4  & $(1.60\pm0.36)\times10^{-3}$ & $1.5\times 10^{-13}$ \\
{\it ROSAT}/HRI    & rh600505a01 & 1995-04-12 &  49819  & 0.1--2.5 & 20.4  & $< 2.3\times10^{-3}$ & $<2.2\times 10^{-13}$ \\
{\it ROSAT}/HRI    & rh500404a01 & 1995-05-08 &  49845  & 0.1--2.5 & 19.1  & $< 1.0\times10^{-3}$ & $<9.4\times 10^{-14}$ \\
{\it ROSAT}/HRI    & rh500403a01 & 1995-05-09 &  49846  & 0.1--2.5 & 31.4  & $< 6.5\times10^{-4}$ & $<6.1\times 10^{-14}$ \\
{\it ROSAT}/HRI    & rh500492n00 & 1997-09-30 &  50721  & 0.1--2.5 & 23.0  & $< 8.9\times10^{-4}$ & $<8.4\times 10^{-14}$ \\
{\it ROSAT}/HRI    & rh500550n00 & 1998-03-21 &  50893  & 0.1--2.5 & 24.2  & $< 6.2\times10^{-4}$ & $<5.8\times 10^{-14}$ \\
\hline
{\it Chandra}/HRC-I    & 2935 & 2002-09-19 &  52536  & 0.08--10 & 1.8  & $< 1.8\times10^{-3}$ & $<4.4\times 10^{-14}$ \\
{\it Chandra}/ACIS-S   & 2950 & 2002-10-13 &  52560  & 0.5--7  & 19.9  & $< 2.3\times10^{-4}$ & $<2.2\times 10^{-15}$ \\
{\it Chandra}/ACIS-I   & 3550 & 2002-11-09 &  52587  & 0.5--7  & 14.6  & $< 5.2\times10^{-4}$ & $<8.4\times 10^{-15}$ \\
{\it Chandra}/ACIS-I   & 3551 & 2003-10-02 &  52914  & 0.5--7  & 14.8  & $< 2.1\times10^{-4}$ & $<3.2\times 10^{-15}$ \\
{\it Chandra}/ACIS-S   & 4747 & 2003-11-17 &  52960  & 0.5--7  &  5.3  & $< 7.5\times10^{-4}$ & $<7.0\times 10^{-15}$ \\
{\it Chandra}/ACIS-S   & 4748 & 2004-02-22 &  53057  & 0.5--7  &  5.1  & $< 6.1\times10^{-4}$ & $<5.9\times 10^{-15}$ \\
{\it Chandra}/ACIS-S   & 4750 & 2004-02-22 &  53057  & 0.5--7  &  4.7  & $< 6.6\times10^{-4}$ & $<6.9\times 10^{-15}$ \\
{\it Chandra}/ACIS-I   & 14676 & 2012-12-17&  56278  & 0.5--7  &  9.8  & $< 4.6\times10^{-4}$ & $<8.2\times 10^{-15}$ \\
{\it Chandra}/ACIS-I   & 15594 & 2012-12-24&  56285  & 0.5--7  &  9.8  & $<
3.1\times10^{-4}$ & $<5.5\times 10^{-15}$

\enddata
\tablecomments{Energy: Energy band used to estimate the photon counts
from each instrument. pn$^{\rm M}$: Medium filter was used. pn$^{\rm T}$: Thin
filter was used. Quiescence: The summed image in 0.3--10 keV is generated for
all {\it Swift}/XRT in the quiescence state (see text). Count rate: The 95.45\%
confidence upper limit in the corresponding energy band is given when the
source is undetected. Flux: The unabsorbed flux (or its $2\sigma$ upper limit)
in 0.5--10 keV is estimated by assuming a power-law model ($nH =
6\times10^{20}$ cm$^{-2}$ and $\Gamma = 1.7$).} \label{obs}
\end{deluxetable*}


\begin{deluxetable*}{cc|ccc|ccccc}
\tabletypesize{\tiny} \tablewidth{0pt} \tablecaption{Spectral fitting results}
\tablehead{\colhead{Observatory} & \colhead{MJD} \vline & \colhead{$nH$} &
\colhead{$\Gamma$} & \colhead{$\chi^2$/dof} \vline & \colhead{$nH$} &
\colhead{$kT$} &
\colhead{$Norm$} & \colhead{$\chi^2$/dof} & \colhead{Flux}  \\
\colhead{} & \colhead{} \vline & \colhead{($10^{21}$ cm$^{-2}$)} & \colhead{} &
\colhead{} \vline & \colhead{($10^{21}$ cm$^{-2}$)} & \colhead{(keV)} &
\colhead{($\times 10^{-3}$)} & \colhead{} &\colhead{}} \startdata

\hline
{\it XMM-Newton}    & 53332 &  $2.5_{-1.9}^{+2.3}$  & $2.2_{-0.8}^{+1.0}$ &  9.8/12  & $0.9_{    }^{+1.4}$ & $0.77_{-0.24}^{+0.65}$ & $30.0_{     }^{+140.9}$ & 11.0/12 & $2.0_{-0.4}^{+0.8}$\\
{\it XMM-Newton}    & 53408 &  $2.7_{-1.0}^{+1.1}$  & $2.4_{-0.4}^{+0.4}$ &  21.3/28 & $0.6_{    }^{+0.6}$ & $0.88_{-0.15}^{+0.19}$ & $22.4_{-12.1}^{+26.2}$  & 25.1/28 & $2.5_{-0.3}^{+0.3}$\\
{\it XMM-Newton}    & 56844 &  $3.3_{-0.3}^{+0.3}$  & $2.3_{-0.1}^{+0.1}$ &  229.7/137 & $0.5_{-0.1}^{+0.1}$ & $1.16_{-0.05}^{+0.05}$ & $17.1_{ -2.4}^{ +3.0}$ & 151.4/137 & $5.9_{-0.2}^{+0.2}$\\
\hline
{\it Swift}         & 55050--55200$^{\sharp}$ &  $2.2_{-1.9}^{+2.2}$  & $2.0_{-0.5}^{+0.6}$ &  19.2/22 & $0.3_{    }^{+1.5}$ & $1.2_{-0.3}^{+0.4}$ & $4.9_{     }^{+10.5}$ & 18.7/22 & $2.0_{-0.4}^{+0.5}$\\
{\it Swift}         & 56800--57100 &  $3.8_{-1.2}^{+1.5}$  &
$2.4_{-0.3}^{+0.4}$ &  16.5/22 & $1.3_{-0.8}^{+0.9}$ & $0.9_{-0.1}^{+0.2}$ &
$17.4_{-8.9}^{+18.6}$ & 15.3/22 & $2.3_{-0.3}^{+0.3}$
\enddata
\tablecomments{Flux: 0.5--10.0 keV absorbed flux calculated with the
disk black-body model in units of $10^{-13}$ erg~cm$^{-2}$~s$^{-1}$. All
errors are in the 90\% confidence level (1.645 $\sigma$). $\sharp$: The
C-statistic is adopted for the spectral fitting.} \label{fitting}
\end{deluxetable*}

\subsection{{\it Swift} observations}

\begin{figure*}
\begin{center}
\includegraphics[scale=0.6]{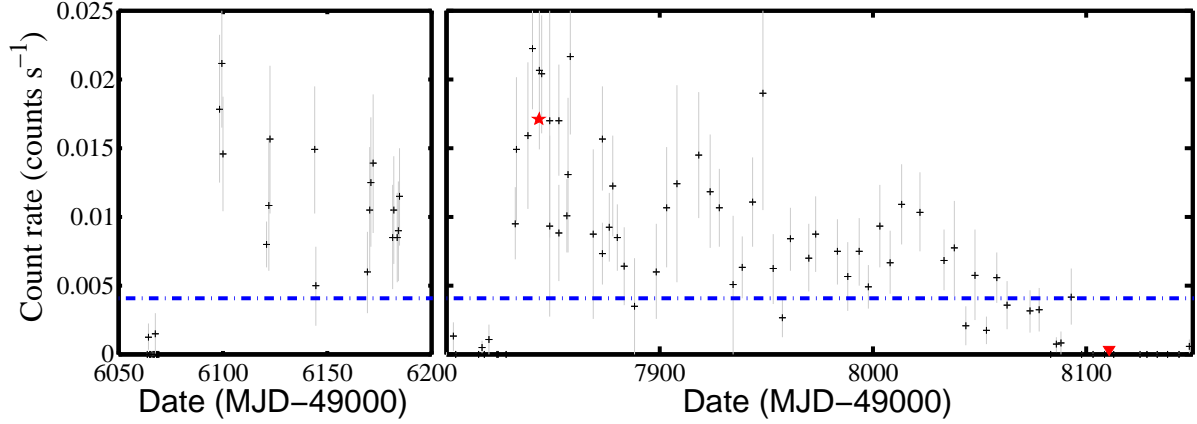}
\caption{{\it Swift}/XRT count rate vs. MJD for two outbursts. The
blue horizontal line mark the {\it Swift}/XRT threshold of $2\sigma$ detection
with an exposure time of 1000 s. The red pentagram and downward-pointing
triangle correspond to the detection and the 2$\sigma$  upper limit derived
from two {\it XMM-Newton} data (converted into XRT count rates).
\label{xrt}}
\end{center}
\end{figure*}

In this work, we analyze all 371 {\it Swift} \citep{gehrels04}
observations made before 2017 July. Two episodic outbursts of 3XMM
J031820.8-663034 have been caught by the {\it Swift} dense observations, and
{\it Swift} monitored the second entire outburst (Figure \ref{xrt}).
The exposure time of individual {\it Swift} observations ranges from
138 s to 7.75 ks, with a mean value of 1.3 ks. We extract the source
counts in 0.3--10 keV band from a circle aperture with a radius of 6 pixels
centered at the source position, and the background from nearby source free
regions. The telescope vignetting and point spread function corrections are
applied by running the {\it Swift} script \texttt{xrtlccorr}.

Due to the limited photons in individual pointings, we stack all
observations during each of the two outbursts. As only 137 photon counts were
collected during the outburst in MJD~55050--55200, we rebin the spectrum to
have at least 5 counts per bin and employ the C-statistic \citep{cash79} in
spectral fitting. During the 2nd outburst, a total of 396 counts were
detected. We group the spectrum to have at least 15 counts per bin and the
common $\chi^{2}$ statistic is applied. The spectral modeling confirms that
the X-ray spectra during the outbursts are very soft, and can be
fitted by either an absorbed steep power-law ($\Gamma \sim 2-2.4$) or a disk
black-body (Table \ref{fitting}). During other {\it Swift}
observations, 3XMM J031820.8-663034 remains undetected. We estimate an upper
limit of count rate $\sim 2.6\times10^{-4}$ cts/s (or an upper limit of L$_X$
$\sim$ $2.8\times10^{37}$ erg/s) by summing all observations in the
off state (with total exposure of $\sim 365$ ks from 285 snapshots).

As shown explicitly in Figure \ref{xrt}, 3XMM J031820.8-663034 has
almost the same peak luminosity during two outbursts. Coincidentally, when the
source reached the peak of the second outburst, the {\it XMM-Newton}
observation recorded an X-ray luminosity of $L_{\rm X} \sim
1.5\times10^{39}$ erg/s (Section \ref{xmm}). Using the {\it XMM-Newton}
spectral fitting model, we convert count rates (and upper limits) from
different instruments into those expected from {\it Swift}/XRT, and plot them
in Figures \ref{xrt} and \ref{lc} for comparison.

\subsection{{\it ROSAT} and {\it Chandra} archival data}

We employ the \textsc{ximage} software to measure the count rate in all 9 {\it
ROSAT}/HRI and 2 {\it ROSAT}/PSPC images \citep{truemper82, pfeffermann87}.
3XMM J031820.8-663034 was detected in 2 images ($> 2\sigma$), and a maximum
count rate ($5.63\times10^{-3}$ cts/s) was obtained on MJD~49527 (see Table 1).
Assuming a power-law with photon index of 1.7, we run the tool
\texttt{WebPIMMS}\footnote{\url{https://heasarc.gsfc.nasa.gov/cgi-bin/Tools/w3pimms/w3pimms.pl}}
and estimate an unabsorbed luminosity of $\sim 1.3\times10^{39}$ erg/s (0.5 --
10 keV), which is consistent with the value reported in \cite{liu05}.

From September, 2002 to December, 2012,  {\it Chandra}
\citep{weisskopf00} visited the source region 9 times. However, the 3XMM
J031820.8-663034 was in quiescence state and was not detected in any
observation. We estimate the flux upper limit with the tasks
\texttt{srcflux/aprates}
\footnote{\url{http://cxc.harvard.edu/ciao/threads/upperlimit/}} in the
\textsc{ciao}\footnote{\url{http://asc.harvard.edu/ciao/}} software (version
4.6.7). The deepest observation (ObsID = 2950) yields the strongest
upper limit to its X-ray luminosity ($\sim 5.6\times10^{36}$ erg/s), which is
more than two orders of magnitude fainter comparing with its maximum luminosity
during the outbursts.

\subsection{Outburst parameters}

The {\it Swift} monitored the entire outburst from MJD $\sim$ 56800 to 57100,
suggesting a fast rise slow decay light curve profile, an outburst duration of
$\sim 240-300$ days and a total energy radiated during the outburst of $\sim
10^{46}$ erg/s. However, we are unable to determine the light curve profile
precisely, including the rise and decay timescales owing to the large
uncertainties of {\it Swift} data. The maximum fluxes recorded in {\it ROSAT},
{\it Swift}, and {\it XMM-Newton} data indicate that the peak luminosity
remains constant ($\sim 1.5\times10^{39}$ erg/s) during different outbursts.
The strongest upper limit for the quiescence state comes from the deepest {\it
Chandra} pointing, indicating a variation amplitude of $> 270$.

\begin{figure*}
\begin{center}
\includegraphics[scale=0.6]{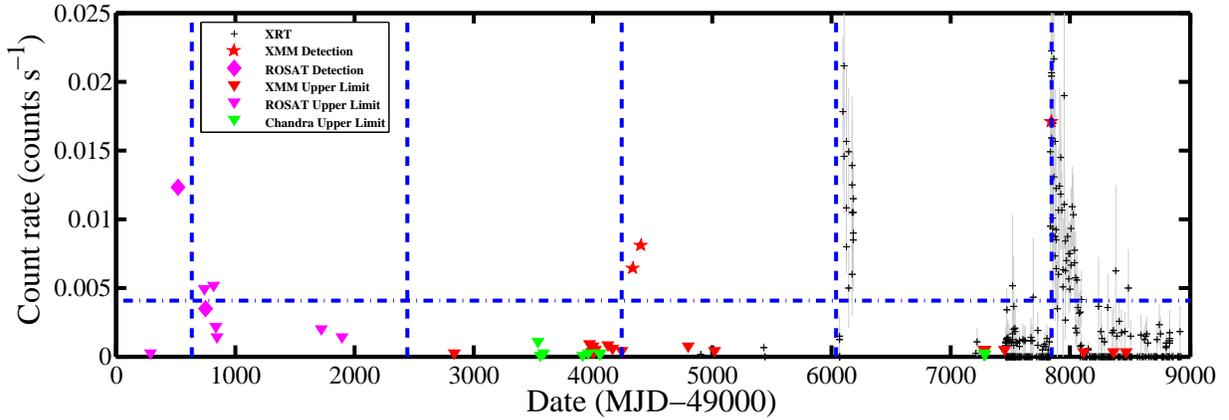}
\caption{X-ray Light curve of 3XMM J031820.8-663034. Count rates and
upper limits from other instruments were converted into $Swift$/XRT count
rates. The blue horizontal line mark the {\it Swift}/XRT threshold of $2\sigma$
detection with an exposure time of 1000 s. The blue vertical lines spaced by
1800 days are plotted to better illustrate the quasi-periodic outbursts.
\label{lc}}
\end{center}
\end{figure*}

We plot X-ray count rates (and upper limits) from all observations in
Fig. \ref{lc}. All count rates from instruments other than {\it Swift}/XRT are
converted into (with \texttt{WebPIMMS}) XRT count rates for comparison. For
three XMM detections, the best-fit spectral models were adopted for the
conversion. For {\it ROSAT} observations and all other upper limits, as
accurate spectral modeling is unavailable, we assume a typical low/hard state
spectrum for conversion, i.e. a hard power-law model ($nH = 6\times10^{20}$
cm$^{-2}$ and $\Gamma = 1.7$).

\begin{figure}
\begin{center}
\includegraphics[scale=0.45]{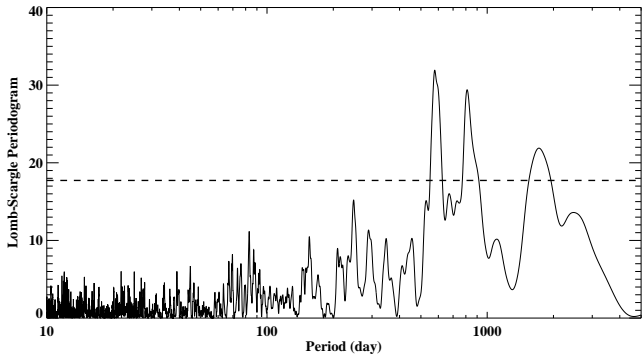}
\caption{Lomb-Scargle periodogram for 3XMM J031820.8-663034. The dashed
line indicates 99.9\% significance.\label{power}}
\end{center}
\end{figure}

\begin{figure}
\begin{center}
\includegraphics[scale=0.5]{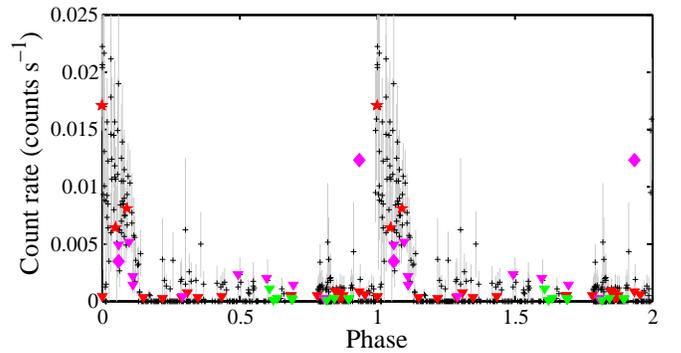}
\caption{Folding the light curve in Figure \ref{lc} over a period of 1800 days.
\label{phase}}
\end{center}
\end{figure}

It seems that the source enters into outburst regularly with a recurrent time
of $\sim 1800$ days. We calculate the Lomb-Scargle periodogram
\citep{horne1986} in a timescale range from 10 days to 5000 days with 50,000
independent frequencies. Since the source was non-detected during most
observations, omitting these non-detections would yield too sparse sampling,
and no periodical signal emerges in the periodogram. Here, we take all data
into account, and 2$\sigma$ upper limits to the count rates are
adopted for those non-detections. As can be seen in Figure \ref{power}, the
signal at $\sim$ 1750 days and its 2nd/3rd harmonic frequencies (P $\sim$
815/580 days) are higher than 99.9\% white noise confidence level. Because the
structured window function might lead to aliasing of signals in the data
\citep[e.g. ][]{vanderPlas17}, we examine the window power spectrum and do not
find any significant feature. We further model the periodogram in the
range of 1500-2000 days with a Gaussian, yielding a best-fit period of 1744
days with a FWHM of $\sim$ 418 days. The recurrent time of $\sim 1800$
days is also verified with the folded light curve (Figure \ref{phase}).
However, we would like to caution that the periodic signal can not be
robustly confirmed for the following reasons: (1) The source was not detected
in many observations, and upper limits used in the test may introduce
large uncertainty. (2) The confidence level of signals would be lower than the
99.9\% with different assumptions of the noise (e.g. red noise).
Future monitoring data are required to check the recurrence period.

\section{Discussion \& Conclusions}

In this work, we investigate the wealth of X-ray data of 3XMM J031820.8-663034
and reveal four periods of activities, making it as a transient ULX.
The outburst properties are summarized as follows: (1) The outbursts
likely occur regularly; (2) One outburst with sufficiently
dense sampling exhibits a fast rise slow decay profile; (3) The peak X-ray
luminosity $L_{\rm peak} \sim 1.5\times10^{39}$ erg/s does not change much
among different outbursts; (4) The source occupies the thermal state at the
high luminosity; (5) The deepest {\it Chandra} observation provides a
luminosity upper limit of $\sim 5.6\times10^{36}$ erg/s for the quiescent
state.

The evolving soft X-ray spectra from 3XMM J031820.8-663034 imply its accretion
nature and the accreting object could be a black hole. High-mass X-ray binaries
commonly consist of a neutron star and a young massive star, and have very hard
X-ray spectra \citep[e.g. ][]{fabbiano06, walter15, wang16b}. Thereby, the
scenario of a canonical high-mass X-ray binary is disfavored for 3XMM
J031820.8-663034. Because its peak luminosity marginally puts the source in the
category of ULXs, below we compare its outburst properties with normal
transient black hole binaries and other known transient ULXs including the
intermediate-mass black hole candidate ESO 243-49 HLX-1.

\subsection{Association with NGC 1313}

Before discussing the nature of outbursts, we need to determine whether 3XMM
J031820.8-663034 is associated with NGC 1313, or a foreground/background
object. The high Galactic latitude ($l = 283.3634\degr$ and $b =
-44.6295\degr$) indicates that the source is unlikely a foreground star.
Moreover, if the source is at a distance of less than 10 kpc, its peak X-ray
luminosity during the outbursts would be less than $10^{34}$ erg/s. The X-ray
spectrum of such a very faint X-ray transient should be dominated by
non-thermal component \citep[e.g. ][]{mcclintock06, weng15, wijnands15},
contradicting the {\it XMM-Newton} and {\it Swift} observations which are in
favor of the thermal dominated spectra. We can therefore exclude the
possibility of a Galactic counterpart.

Alternatively, we estimate the probability of background QSO/AGN using X-ray
log$N$--log$S$ \citep[e.g. ][]{wang16a}. According to the new {\it XMM-Newton}
detection, the source has an absorbed flux in 0.5--2 keV of $2.3\times10^{-13}$
erg/cm$^{2}$/s. We would expect $\sim$ 0.5 contaminating source per square
degree with the same or higher flux based on the Lockman Hole log$N$--log$S$
relation \citep{hasinger98, mushotzky00, moretti03, wang16a}. Meanwhile, the
$D_{25}$ isophote of NGC 1313 is about $\sim 9.1\arcmin$ \citep{liu05}, and the
distance between 3XMM J031820.8-663034 and the center of NGC 1313 is $\sim
0.8\arcmin$. Therefore, the number of expected background sources within the
D25 of NGC 1313 (or in 0.8\arcmin) is only $\sim 0.01$ (or $9\times10^{-5}$).
Furthermore, the chance that a background source has similar dramatic
variations is even lower. We thus conclude the association of 3XMM
J031820.8-663034 with NGC 1313 is convincing.

\begin{deluxetable}{lcc}
\tabletypesize{\tiny} \tablewidth{0pt} \tablecaption{Outburst
parameters} \tablehead{\colhead{} & \colhead{3XMM J031820.8-663034}  &
\colhead{HLX-1$^\sharp$}} \startdata \hline
$T_{\rm recurrent}$ & $\sim 1800$ days  & $\sim 400$ days \\
Duration & $\sim 240-300$ days & $\sim 113$ days \\
Duty cycle & $\sim 0.15$ & $\sim 0.3$   \\
$L_{\rm peak}$   & $\sim 1.5\times10^{39}$ erg/s  & $\sim 1.2\times10^{42}$ erg/s \\
Fluence$^\dag$  & $\sim 10^{46}$ erg  & $\sim 5.8\times10^{48}$ erg   \\
Amplitude  & $> 270$  & $\sim 20-50$
\enddata

\tablecomments{$\sharp$: Averaged values are listed for HLX-1
\citep{godet14, yan15}. $\dag$: Fluence is referred to the total energy
radiated during one outburst. \label{log}}
\end{deluxetable}

\subsection{Repeated outbursts}

A straightfoward explanation for the regular outbursts is that the mass
transfer rate is significantly enhanced during the periastron passage of the
donor star bounded to the black hole in an eccentric orbit. However
the formation of a black hole binary with such long period ($\sim
1800$ days) could be challenging. To our knowledge, GRS~1915+105 has
the longest orbital period \citep[$\sim 33.5$ days, ][]{greiner01} among LMXBs
\citep{liu07}. Meanwhile, very few high-mass X-ray binaries have periods longer
than 10$^{3}$ days \citep{liu06}, e.g. PSR B1259-63 \citep[$\sim 3.4$ years,
][]{johnston94} and PSR J2032+4127 \citep[$\sim 25-50$ years, ][]{lyne15,
ho17}, both of which are faint in X-ray ($L_{\rm X} < 10^{35}$ erg/s). On the
other hand, during the early observations HLX-1 displayed regular outbursts at
an interval of $\sim 1$ yr \citep{farrell09a, webb12}. However, the initial
hypothesis that the recurrence time corresponded to the binary period became
controversial due to the increasing of the detected recurrence time
\citep{godet14, weng18}.

In addition, the super-orbital periods have been reported in tens of X-ray
binaries \citep{sood07, farrell09b}, and some of them could be more than
$10^{3}$ days \citep[e.g. $\sim$  1667 days for LS I +61$\degr$ 303, ][]{li12},
similar to the recurrent time of 3XMM J031820.8-663034. However the amplitude
of super-orbital modulation \citep[e.g. ][]{smith02, corbet13} is significantly
smaller than that of 3XMM J031820.8-663034 ($> 270$). Thus, whether the
outbursts of 3XMM J031820.8-663034 are some kind of super-orbital modulation is
still questionable.

Note a small number of LMXBs, e.g. 4U~1630-47 \citep{parmar95,
capitanio15} and H1743-322 \citep{yan15}, exhibited some successive (not all)
outbursts equally spaced in time. The light curve profiles of these outbursts
could be complicated, but not always the typical ``fast-rise
exponential-decay''. It was suggested that the periodicity was
resulted from the DIM, or sometimes with additional perturbation from the
companion star mass transfer \citep{capitanio15}. Here, since only four
outbursts are recorded for 3XMM J031820.8-663034 and the observation cadence is
incomplete, we suggest that more monitoring data are required to check whether
the quasi-periodic behaviors are temporary or represent some process in
physics. As further discussed below, all available data at the current
stage can be interpreted with the DIM.


\subsection{Outburst mechanism}

There are two transient ULXs detected in M31, i.e. CXOM31~J004253.1+411422
\citep{kaur12, middleton12} and XMMU~J004243.61+412519 \citep{middleton13}.
\cite{middleton12} found the X-ray luminosity of CXOM31~J004253.1+411422
steadily declined from $5\times10^{39}$ erg/s to $6\times10^{38}$ erg/s over
1.5 month \citep[see also ][]{kaur12}. XMMU~J004243.61+412519 entered
an outburst in 2012 January, reached a peak X-ray luminosity of $\sim
1.26\times10^{39}$ erg/s within a few days, and then decayed slowly
\citep{middleton13}. Their outburst parameters resemble those found in
Galactic LMXBs, indicating a DIM origin for these two sources as well
\citep[e.g. ][]{yan15a}. However only one outburst had been observed
for each source thus the recurrence timescale is unavailable.

\begin{figure}
\begin{center}
\includegraphics[scale=0.45]{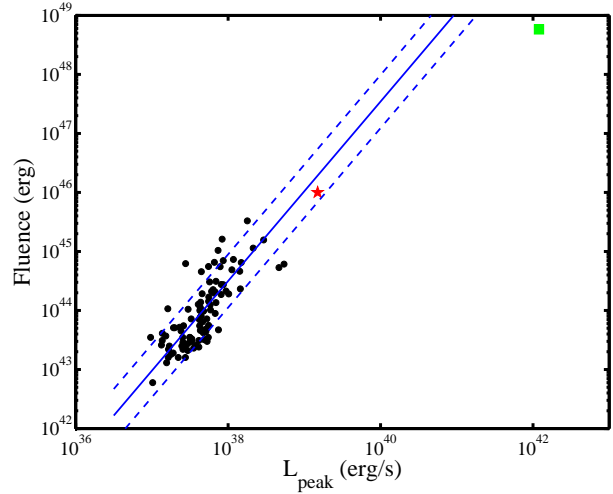}
\caption{Total energy radiated during the outburst vs. peak X-ray luminosity.
The filled circles correspond to the outbursts of LMXBs adopted from
\cite{yan15a}. The blue lines indicate the best-fit result with a linear model
in a logarithmic scale (solid) and the 3$\sigma$ confidence intervals (dashed).
The red pentagram and the green square mark 3XMM J031820.8-663034 and HLX-1,
respectively. \label{lpeak_e}}
\end{center}
\end{figure}

It is worth to note that the recurrence outbursts of the best studied transient
ULX, HLX-1, are analogous to the outbursts of 3XMM J031820.8-663034 in many
aspects. Meanwhile, the outbursts of HLX-1 have significantly higher luminosity
at both active and quiescent states, and smaller amplitude of luminosity
variation. The total energy radiated during an outburst is about 2-3 orders of
magnitude larger than the averaged value observed in 3XMM J031820.8-663034
(Table \ref{log}). \cite{yan15} argued that HLX-1 and LMXBs follow the same
linear relationship between the hard-to-soft state transition luminosity and
the peak luminosity, but the data of HLX-1 deviate the correlation between the
X-ray fluence and peak luminosity observed in LMXBs (Figure \ref{lpeak_e}). The
minimum observed X-ray luminosity of HLX-1 is up to $\sim 3\times10^{40}$
erg/s, indicating that the thin disk within a very large radius ($\sim 10^{13}$
cm) is fully ionized because of large accretion rate and the additional heating
by the strong irradiation \citep{dubus01, soria17}. In this case, the local
thermal-viscous instability should be ignited (if existed) at even larger
physical radius (not the black hole mass scaled radius). Such instability would
take at least 100 years to reach the inner most accretion disc, which however
contradicts the observed value of $\sim 100$ days \citep{lasota11}. Therefore,
the HLX-1 outbursts are unlikely due to thermal-viscous instability (i.e. DIM)
but might be attributed to the radiation pressure instability \citep{lasota11,
sun16} or other instabilities.

Investigating the optical/UV and X-ray data, \cite{soria17} proposed an
oscillating wind scenario for HLX-1: currently its accretion rate is
on average of a few percent Eddington and its accretion disk is quite large
($\sim 10^{13}$ cm); but only the inner region ($\leq 10^{12}$ cm) of the
inflow is modulated by the wind instability \citep{begelman83,
shields86}, which drives the outbursts at a timescale of $\sim 1$ yr.
Such model might also work for V404~Cyg \citep{munoz16}. Due to the limited
data, we do not know whether the strong wind can be launched from 3XMM
J031820.8-663034, and whether the wind instability model is applicable to its
repeated outbursts. The model is presented here as an option. Alternatively, we
argue that the outburst properties of 3XMM J031820.8-663034 reported in this
paper can be understood in the framework of DIM.

The X-ray luminosity at the quiescent state of 3XMM J031820.8-663034
($< 5.6\times10^{36}$ erg/s) indicates that the disk is cool due to a low
accretion rate and weak irradiation. The partial hydrogen ionization
instability (i.e. thermal-viscous instability) thus emerges at a much smaller
radius and allows the DIM to work. Additionally, we point out that most of the
outbursts properties are consistent with the expectation of DIM: (1)
The light curve shows a fast rise slow decay profile. (2) It falls on
the same relation between the outburst fluence and the peak luminosity that was
found for LMXBs \citep{yan15a}. Such correlation is expected in DIM since the
outburst peak increases with the mass of accretion disk \citep[e.g.
][]{dubus01}. Assuming a constant radiative efficiency, the mass of accretion
disk and the peak accretion rate can be estimated from the fluence and the
$L_{\rm peak}$, respectively. (3) Taking the truncation and irradiation effects
into account, theoretical DIM models yield the recurrence time of 1
$\sim$ 180 years \citep{dubus01, lasota01}, conforming to that of 3XMM
J031820.8-663034 ($\sim$ 1800 days).

In addition to X-ray, the nature of the outbursts can be explored with
multi-wavelength data. The optical/UV data could provide key information of the
companion star and the X-ray irradiated accretion disk \citep[e.g.
][]{rykoff07, weng15, soria17}. Furthermore, the connections between accretion
flows and the radio jet have been widely studied \citep[e.g. ][]{fernder04,
zhang14}. A stable jet is commonly detected in the low/hard state, while the
discrete ejection events are found to be associated with the transitions
between the low/hard and the high/soft states. We search the radio images in
the literature and the
\texttt{SkyView}\footnote{\url{https://skyview.gsfc.nasa.gov/current/cgi/query.pl}},
but do not find the point like source at the position of 3XMM J031820.8-663034
in the 1.4 GHz radio continuum map \citep{ryder93} nor the SUMSS 843 MHz image.
A detailed analysis on these data is beyond the purpose of this paper.

\subsection{Accretion state \& black hole mass}

Most Galactic LMXBs are in the regime of sub-Eddington accretion with variable
X-ray emission. Although different state classifications have been proposed by
different authors \citep[e.g. ][]{mcclintock06, zhang13, yuan14}, the low/hard
and the high/soft states are normally undisputed. The low/hard state is
characterized by a hard spectrum and strong rapid variations. In contrast, the
high/soft state is dominated by a thermal disk component and has a low level of
variability. Meanwhile, several Galactic LMXBs can occasionally be brighter
than $10^{39}$ erg/s, e.g. GRS~1915+105 \citep[e.g. ][]{belloni00, yan17},
V4641~Sgr \citep{revnivtsev02} and V404~Cyg \citep[e.g. ][ and references
therein]{motta17}. These three sources are highly variable on time-scales of
minutes to hours during the outbursts. In particular, during the 2015 outburst,
V404~Cyg showed violent variations in both X-ray and optical bands
\citep{kimura16} and the non-thermal dominated X-ray emission \citep[e.g.
][]{sanchez17, motta17}. The source did not enter into the canonical high/soft
state, but might be accreting at super-Eddington accretion rate. Its evolution
pattern is distinct from other typical LMXBs.

The new {\it XMM-Newton} detection of 3XMM J031820.8-663034 performed at the
peak of an outburst indicates that its spectrum is dominated by a thermal disk
component and no significant variation is detected within the XMM exposure. The
X-ray properties are consistent with the definition of the high/soft state,
that is, the source reaches a luminosity of $(0.1-1)~L_{\rm Edd}$ during the
outbursts. Compared to those Galactic LMXBs at the high/soft state, 3XMM
J031820.8-663034 has a higher peak X-ray luminosity ($\sim 1.5\times10^{39}$
erg/s), which might indicate a heavier black hole (tens of solar masses) hosted
in the system.  The MsBH scenario is also supported by the fitted disk
black-body normalization \citep[$\sim 0.02-0.03$, ][]{arnaud96}, which
corresponds to a radius of $\sim 300$ km with a inclination angle of 60\degr\
and a spectral hardening factor of 1.7 adopted. If the accretion disk extends
to the innermost stable circular orbit, the derived radius corresponds to $\sim
30-200 M_{\odot}$ for a Schwarzschild and a maximally rotating black hole,
respectively.

\acknowledgements {We thank the anonymous referee for the helpful comments. We
acknowledge the use of public data from the High Energy Astrophysics Science
Archive Research Center Online Service. We thank Drs. Wei-Min Gu, Zhen-Yi Cai
and C.-Y. Ng for many valuable discussions. This work is supported by the
National Natural Science Foundation of China under grants 11703014, 11673013,
11421303, and 11573023, the Natural Science Foundation from Jiangsu Province of
China (Grant No. BK20171028), and the University Science Research Project of
Jiangsu Province (17KJB160002). J.X.W. thanks support from Chinese Top-notch
Young Talents Program, and CAS Frontier Science Key Research Program
QYCDJ-SSW-SLH006.}

{\it Facilities: ROSAT, XMM-Newton, Chandra,} Neil Gehrels {\it Swift}
Observatory.

{\it Software:} \textsc{sas}, \textsc{heasoft}, \textsc{ciao}.

\end{document}